# 16-QAM Quantum Receiver with Hybrid Structure Outperforming the Standard Quantum Limit

Yuan Zuo[1], Ke Li[2], and Bing Zhu[1]

[1]*Department of Electronic Engineering and Information Science, University of Science and Technology of China, China*
[2]*Nanjing Research Institute of Electronic Technology, China*
zuoyuan@mail.ustc.edu.cn

**Abstract:** We present a quantum receiver for 16-QAM signals discrimination with hybrid structure containing a homodyne receiver and a displacement receiver, which can outperform the SQL, and the performance can be improved by an optimized displacement.

## 1. Introduction

It is known for decades that quantum receivers can discriminate different signals below the standard quantum limit (SQL) and achieve even lower error probability limit called Helstrom bound [1]. For phase shift keying (PSK) and pulse position modulation (PPM) signal discrimination, many kinds of quantum receivers have been proposed theoretically and some of them demonstrated experimentally [2-11]. However, there are few literatures paying attention to QAM signals discrimination. Inspired by Bondurant [4] and Müller [6, 12], we present an original quantum receiver with a hybrid structure, which contains a homodyne receiver and a displacement receiver. Numerical simulation has shown that it can outperform the SQL. What's more, its performance can be even better, if we optimize the displacement, especially when the signals are weak.

## 2. Schemes and Performances

Fig. 1 (*a*) shows the phase space configuration of 16-QAM signals and our receiver scheme. In order to discriminate the 16-QAM signals below SQL, the incoming signal is first split into two equivalent parts by a beam splitter (BS). Then one beam is directly fed into a homodyne detector (HD), whilst the other one is detected using a displacement receiver, which is controlled using a feed-forward and feedback strategy as shown in Fig. 1 (*a*). The HD determines the P quadrature of the incoming signal. After HD, 16 hypotheses are reduced to only 4 hypotheses $H_{1,2,3,4}$ as shown in Fig. 1 (*b*). Next we use a displacement receiver to discriminate these four signals. The displacement receiver uses a displacement operator and an ideal single photon detector (SPD). It sequentially nulls signals $1 \to 2 \to 3 \to 4$ just like Bondurant receiver [4]. At first, the receiver null signal 1 until a photon counting event happens. Every time a photon counting event happens, the receiver immediately changes the displacement to null the next signal and keeps counting. Suppose $N$ is the photon count number in whole symbol interval, if $N \leq 3$, we choose the hypothesis $H_{N+1}$; if $N > 3$, choose $H_4$. However, exact nulling is not a good idea when signal is weak [6, 12]. By adding an additional optimal displacement $\beta$, as shown in Fig. 1(*b*), the error probability will be further reduced.

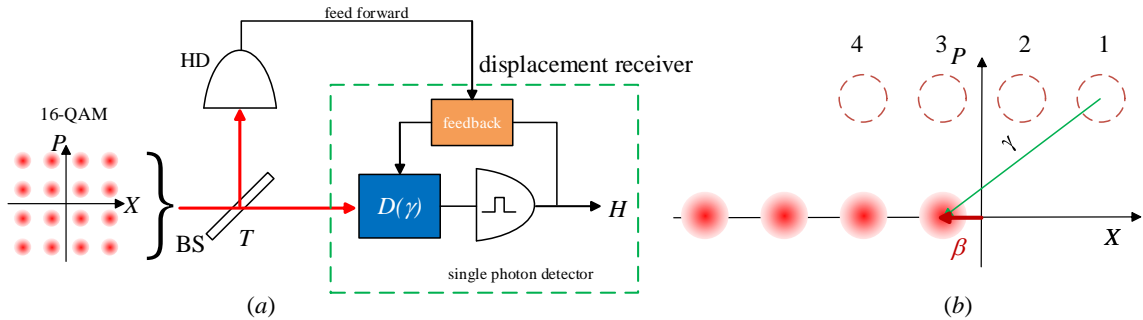

Fig. 1. (*a*) Hybrid structure for 16-QAM signals. The incoming signal is equally divided into two parts by beam splitter (BS). One part is detected by a homodyne detector (HD) and the result is feed forward to the displacement receiver. (*b*) Displacement strategy for displacement receiver. When using the exactly nulling, choose $\beta = 0$. When using the optimal displacement, choose $\beta$ through numerical optimization.

We simulate our receiver under two different configurations, exact nulling (Type I) and optimal displacement (Type II). Results are shown in Fig. 2. As in Fig. 2 (*a*), the optimal displacement decays with the average photon number increasing. When the signal is strong enough, the additional displacement $\beta$ can be ignored, so the performance of the Type I receiver approaches to the performance of the type II receiver as shown in Fig. 2(*b*). In this case, type I receiver (blue) and type II receiver (cyan) can both outperform SQL (red). However, when the signal is weak, the performance of the type I is above the SQL, but the type II receiver can still outperform SQL within a wider range of average photon number.

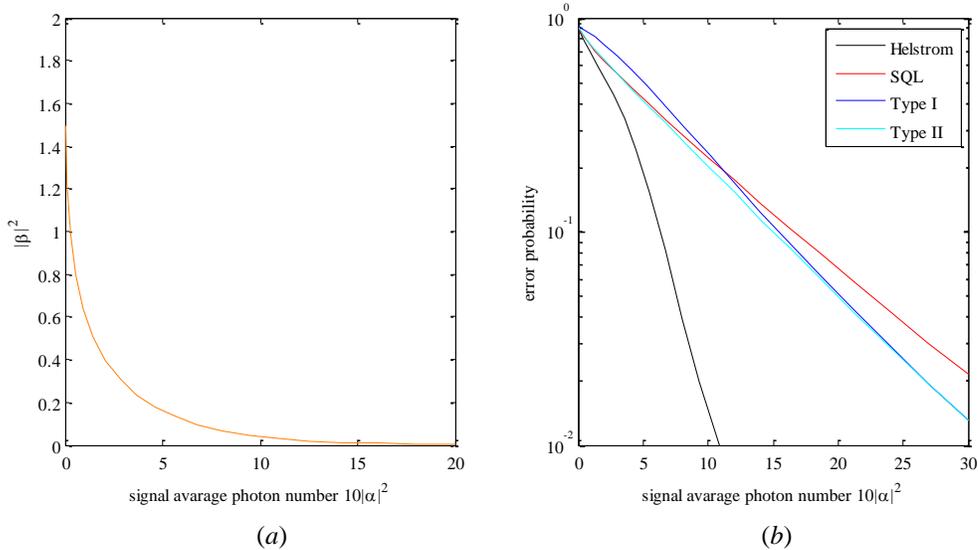

Fig.2. (*a*) Optimal displacement parameter $|\beta|^2$ as a function of signal energy. (*b*) Error probability for hybrid structure receiver using exact nulling (type I, blue) and optimal displacement (type II, cyan) compare to SQL (red) and Helstrom limit (black).

## 3. Conclusion

We have proposed a receiver with hybrid structure for 16QAM signal discrimination, and ran numerical simulation to verify that it can outperform SQL in bright light regime. Its performance can be further improved by using optimal displacement, especially when the signal is weak.